\documentclass[preprint,showpacs,preprintnumbers,amsmath,amssymb]{revtex4}

\usepackage{slashed,color,graphicx}
\usepackage{hyperref}

\hypersetup{
colorlinks = true,
linkcolor = blue,
citecolor = magenta
}
\begin{document}


\title{XENON1T excess in local $Z_2$ DM models with light dark sector}


\author{Seungwon Baek}
\email[]{sbaek1560@gmail.com}
\affiliation{Department of Physics, Korea University, Seoul 02841, Korea}

\author{Jongkuk Kim}
\email[]{jongkuk.kim927@gmail.com}
\affiliation{School of Physics, KIAS, Seoul 02455, Korea}

\author{P. Ko}
\email[]{pko@kias.re.kr}
\affiliation{School of Physics, KIAS, Seoul 02455, Korea}


\preprint{KIAS-P20032}
\date{\today}

\begin{abstract}
Recently XENON1T Collaboration announced that they observed some excess  in the electron recoil 
energy around a 2--3 keV.  We show that this excess can be interpreted as exothermic scattering of 
excited dark matter 
on atomic electron through dark photon exchange.  
We consider DM models with local dark $U(1)$ gauge symmetry that is spontaneously broken into 
its $Z_2$ subgroup by Krauss-Wilczek mechanism. In order to explain the XENON1T excess with 
the correct DM thermal relic density within freeze-out scenario, all the particles in the dark sector 
should be light enough, namely $\sim O(100)$ MeV for scalar DM and $\sim O(1-10)$ MeV for fermion 
DM cases. And even lighter dark Higgs $\phi$ plays an important role in the DM relic density calculation: 
$X X^\dagger \rightarrow Z' \phi$ for scalar DM ($X$)  and $\chi \bar{\chi} \rightarrow \phi \phi$ 
for fermion DM ($\chi$) assuming $m_{Z'} > m_\chi$.  
Both of them are in the $p$-wave annihilation, and one can easily evade 
stringent bounds from Planck data on CMB on the $s$-wave annihilations, assuming other dangerous 
$s$-wave annihilations are kinematically forbidden.  
 \end{abstract}


\maketitle

\section{Introduction}

Recently XENON 1T Collaboration reported that they found electron recoil excess around 2-3 keV 
with 3.5 $\sigma$ significance analyzing data for an exposer of 0.65 ton-year \cite{Aprile:2020tmw}.  
There is an issue on the tritium contamination to be resolved.  
This energy region is sensitive to solar axion search, but the interpretation of this excess in terms 
of solar axion is in conflict with astrophysical bounds on the axion coupling to electron.
XENON1T Collaboration also interprets the excess in the context of magnetic moment of solar 
neutrino and absorption of light bosonic dark matter \cite{Aprile:2020tmw}. 
After the accouncement of XENON1T Collaboration, there appeared a number of papers that 
address various issues related with this excess 
\cite{Takahashi:2020bpq,OHare:2020wum,Kannike:2020agf,Alonso-Alvarez:2020cdv,Fornal:2020npv,Amaral:2020tga,Boehm:2020ltd,Harigaya:2020ckz,Bally:2020yid,Su:2020zny,Du:2020ybt,DiLuzio:2020jjp,Dey:2020sai,Chen:2020gcl,Bell:2020bes,AristizabalSierra:2020edu,Buch:2020mrg,Choi:2020udy,Paz:2020pbc,Lee:2020wmh,Cao:2020bwd,Robinson:2020gfu,Khan:2020vaf,Primulando:2020rdk,Nakayama:2020ikz,Gelmini:2020xir,Jho:2020sku,Bramante:2020zos,Baryakhtar:2020rwy,An:2020bxd,Zu:2020idx,Gao:2020wer,Budnik:2020nwz,Lindner:2020kko,Bloch:2020uzh,Chala:2020pbn,DeRocco:2020xdt,Dent:2020jhf,McKeen:2020vpf,
Coloma:2020voz,An:2020tcg,DelleRose:2020pbh,Dessert:2020vxy,Bhattacherjee:2020qmv,Ge:2020jfn,Chao:2020yro,Gao:2020wfr,Ko:2020gdg,Hryczuk:2020jhi}.

In this paper, we interpret this electron recoil excess in terms of exothermic DM scattering on atomic 
electron bound to Xe in the inelastic DM models.   We shall consider both complex scalar 
\cite{Baek:2014kna} and Dirac fermion DM models \cite{Ko:2019wxq} with local $U(1)$ dark gauge 
symmetry which is spontaneously broken into its $Z_2$ subgroup by Krauss-Wilczek mechanism 
\cite{Krauss:1988zc}.  In this framework, the mass difference ($\delta$)  
between the DM and the excited DM (XDM) is generated by dark Higgs mechanism, 
and there is no explicit violation of local gauge symmetry related with the presence of dark photon.  
On the other hand,  in a number of literature, the mass difference $\delta$ is often introduced 
by hand in terms of dim-2 (3) operators for scalar (fermion) DM.  
Then local gauge symmetry is broken explicitly and softly.  
Introducing dark gauge boson (or dark photon) in such a case would be theoretically inconsistent, 
since the current dark gauge fields couple is not a conserved current.  
There will appear some channels where high energy behavior of the scattering amplitudes violate
perturbative unitarity, in a similar way with the $W_L W_L \rightarrow W_L W_L$ scattering violates 
unitarity if $W$ boson mass is put in by hand.  One of the present authors  pointed out this issue 
in fermionic DM model (see Appendix A of Ref.~\cite{Ko:2019wxq}). 

Local $Z_2$ scalar \cite{Baek:2014kna}  and fermion DM models \cite{Ko:2019wxq}
have been studied by authors for the $O(100)$ GeV -- O(1) TeV WIMP scenarios.  
In this paper we explore the same models for light DM mass  $\lesssim O(1)$ GeV in order to 
evade the strong bounds from direct detections experiments searching for signals of DM scattering 
on various nuclei. In particular we will emphasize that the DM thermal relic density and the XENON1T  
electron recoil excess with a few keV could be simultaneously accommodated if dark Higgs boson 
is light enough that ${\rm DM} + {\rm XDM} \rightarrow Z^{' *} \rightarrow Z' \phi$ is kinematically 
open.  This channel will play an important role when $m_{DM} < m_{Z'}$, as we  shall demonstrate 
in the following.  In order to explain the XENON1T excess in terms of 
${\rm XDM} + e_{atomic} \rightarrow {\rm DM} + e_{free}$ with a kinetic mixing, both dark photon and
(X)DM mass should be sub-GeV, more specifically $\sim O(100)$ MeV, in order to avoid the stringent 
bounds on the kinetic mixing parameter. For such a light DM, one has to consider the DM annihilation 
should be mainly in $p$-wave, and not in $s$-wave,  in order to avoid strong constraints from CMB 
(see \cite{Slatyer:2015jla,Leane:2018kjk} and references therein).  

For this purpose it is crucial to have dark Higgs $(\phi)$, since they can play a key roles in the $p$-wave 
annihilations of DM at freeze-out epoch:
\begin{eqnarray*}
X X^\dagger & \rightarrow & Z^{' *} \rightarrow Z^{'} \phi  , \\
\chi \overline{\chi} & \rightarrow & \phi \phi , 
\end{eqnarray*}
where $X$ and $\chi$ are complex scalar and Dirac fermion DM, respectively. 
At freeze-out epoch, the mass gap is too small ($\Delta m \ll T$) and we can consider DM as 
complex scalar or Dirac fermion.  In the present Universe,  we have $T \ll \Delta m$ and so we have 
to work in the two component DM picture for XENON1T electron recoil.  
It can not be emphasized enough that these channels would not be possible without dark Higgs 
$\phi$, and  it would be difficult to make the DM pair annihilation be dominated by the $p$-wave 
annihilation.


\section{Models for (excited) DM}
\subsection{Scalar DM model}
The dark sector  has a gauged $U(1)_X$ symmetry. There are two scalar particles in the dark sector $X$ and $\phi$ with
$U(1)_X$ charges 1 and 2, respectively. They are neutral under the SM gauge group.
After $\phi$ gets VEV, $\langle \phi \rangle = v_\phi/\sqrt{2}$, the gauge symmetry is spontaneously broken down to discrete $Z_2$.
The  $Z_2$-odd $X$  becomes the DM candidate.
The model Lagrangian is in the form~\cite{Baek:2014kna}
\begin{eqnarray}
{\cal L} & = & {\cal L}_{\rm SM}  
- \frac{1}{4} \hat{X}_{\mu\nu} \hat{X}^{\mu\nu} - \frac{1}{2} \sin \epsilon \hat{X}_{\mu\nu} \hat{B}^{\mu\nu} 
+ D^\mu \phi^\dagger D_\mu \phi + D^\mu X^\dagger D_\mu X 
 - m_X^2 X^\dagger X + m_\phi^2 \phi^\dag \phi
 \nonumber \\
 &&  
- \lambda_\phi \left( \phi^\dagger \phi \right)^2 - \lambda_X \left( X^\dagger X \right)^2\
- \lambda_{\phi X} X^\dagger X \phi^\dagger \phi
- \lambda_{\phi H} \phi^\dagger \phi H^\dagger H
- \lambda_{HX} X^\dagger X H^\dagger H  
\nonumber\\
&&-  \mu \left( X^2 \phi^\dagger +  H.c. \right), 
\label{eq:model}
\end{eqnarray}
where $\hat{X}_{\mu\nu}$ ($B_{\mu\nu}$) is the field strength tensors of $U(1)_X$ ($U(1)_Y$) gauge boson in the interaction basis. 

We decompose the $X$ as
\begin{align}
X ={1 \over \sqrt{2}} (X_R + i X_I),
\end{align}
and $H$ and $\phi$ as
\begin{align}
H = 
\begin{pmatrix}
0 \\
{1 \over \sqrt{2}} (v_H + h_H)
\end{pmatrix}, \quad
\phi = {1 \over \sqrt{2}} (v_\phi + h_\phi),
\label{eq:ScalarDecomp}
\end{align}
in the unitary gauge.  

The dark photon mass is given by
\begin{align}
m_{Z'}^2 \simeq (2 g_X v_\phi)^2,
\end{align}
where we neglected the corrections from the kinetic mixing, which is second order in $\epsilon$ parameter.
The masses of $X_R$ and $X_I$ are obtained to be
\begin{align}
m_R^2 &= m_X^2+{1 \over 2} \lambda_{HX} v_H^2 + {1 \over 2} \lambda_{\phi X} v_\phi^2 + {\mu \over \sqrt{2}} v_\phi, \nonumber\\
m_I^2 &= m_X^2+{1 \over 2} \lambda_{HX} v_H^2 + {1 \over 2} \lambda_{\phi X} v_\phi^2 - {\mu \over \sqrt{2}} v_\phi,
\end{align}
and the mass difference, $\delta \equiv m_R - m_I \simeq \mu v_\phi/\sqrt{2} m_X$.
Since the original $U(1)_X$ symmetry is restored by taking $\mu=0$, small $\mu$ does not give rise to fine-tuning problem.
The mass spectrum of the scalar Higgs sector can be calculated by diagonalising the mass-squared matrix
\begin{align}
\begin{pmatrix}
2 \lambda_H v_H^2 & \lambda_{\phi H} v_H v_\phi \\
\lambda_{\phi H} v_H v_\phi &2 \lambda_\phi v_\phi^2 
\end{pmatrix},
\label{eq:H-Mass}
\end{align}
which is obtained in the $(h_H, h_\phi)$ basis.
We denote the mixing angle to be $\alpha_H$ and the mass eigenstates to be $(H_1,H_2)$, 
where $H_1$ is the SM Higgs-like state and $H_2 (\equiv \phi)$ is mostly dark Higgs boson.
Since we work in the small $\alpha_H$ in this paper, the VEV of $\phi$ is approximated to be,
$v_\phi \simeq {m_{H_2} / \sqrt{2 \lambda_\phi}}$, while $\alpha_H \simeq  \lambda_{\phi H} v_\phi / 2 \lambda_H v_H$.  
 
The mass eigenstates $Z_\mu$ and $Z'_\mu$ of the neutral gauge bosons can be obtained using the procedure shown in Ref.~\cite{Babu:1997st}.
In the linear order approximation in $\epsilon$ we can write the covariant derivative as
\begin{align}
D_\mu \simeq \partial_\mu + i e Q_{\rm em} A_\mu + i \Big( g_Z (T^3 -Q_{\rm em} s_W^2) + \epsilon g_X Q_X s_W \Big) Z_\mu 
+ i \Big( g_X Q_X - \epsilon e Q_{\rm em} c_W \Big) Z'_\mu, 
\end{align}
where $Q_{\rm em}$ ($Q_X$) is the electric ($U(1)_X$) charge and $A_\mu$ is the photon field.
We note that $Z'$ couples  to the electric charge but not to the weak isospin component $T^3$.
For example, $Z'$ does not couple to neutrinos at this order of $\epsilon$.

To evade the bound from the DM scattering off the nuclei we are considering sub GeV scale DM.
To calculate the relic abundance of the DM we take the $X$ as the physical state  with mass $m_X$ instead of $X_I$ and $X_R$,
{\it i.e.} $m_X \simeq m_R \simeq m_I$,
because the mass difference $\delta$ is much smaller than the freezeout temperature 
$T_f \sim m_X/10$~\cite{Harigaya:2020ckz}.
For this light DM the CMB constraint rules out the $s$-wave annihilation of the DM pair.
So the contribution to the DM annihilation should start from $p$-wave.
We suppress the $X X^\dagger \to Z' Z'$ by choosing $m_{Z'} > m_X$.
We also make $X X^\dagger \to H_2 H_2$ subdominant.
To achieve this we suppress the direct coupling of the DM to $\phi$ and $H$ by taking
small $\lambda_{\phi X}$ and $\lambda_{HX}$.
We also take small $\lambda_{\phi H}$ to evade the bound from the Higgs invisible decay.
However, the coupling $\lambda_{\phi H}$ should not be too small to make the DM in thermal contact with the SM plasma. 
Too small $\lambda_{\phi H}$ also makes the $H_2$ lifetime too long, causing cosmological problems.
For example, $H_2$ with mass $\sim 1$ GeV decays dominantly into muon (or strange quark) pairs through mixing $\lambda_{\phi H}$,
whose decay with is given by

\begin{align}
\Gamma(H_2 \to \mu^+ \mu^-) &\simeq \frac{\alpha_H^2}{8 \pi} m_{H_2} \left(m_\mu \over v_H \right)^2 \left(1 -{4 m_\mu^2 \over m_{H_2}^2} \right)^{3/2} \nonumber\\
&\approx \left( {1.1 \times 10^{16}  \, {\rm sec}^{-1}} \right) \alpha_H^2 \left(m_{H_2} \over 1\, {\rm GeV}\right).
\label{eq:Gmm}
\end{align}
We require that $H_2$ lives shorter than 1 sec to evade the constraints from Big Bang nucleosynthesis  (BBN) 
 and the mixing angle is small. For $m_{H_2}=1$ GeV it is translated into $\alpha_H > 9.5 \times 10^{-9}$.
When the muon channel is kinematically forbidden but $m_{H_2} > 2 m_e$, it decays into electron-positron pair. The decay rate
is obtained by replacing $m_\mu$ by $m_e$ in (\ref{eq:Gmm}). For example, for $m_{H_2} = 10$ MeV, we
need $\alpha_H > 2.0 \times 10^{-5}$.
When $m_{H_2} < 2 m_e$, the scalar particle decays into two photons.
In this paper we consider $m_{H_2} \gtrsim 2 m_e$. 
The small mixing parameters are also technically natural by extending the Poincar\'e symmetry~\cite{Foot:2013hna,Baek:2019wdn}. 

The $Z'$ can also decay into charged SM particles through mixing $\epsilon$. 
The decay width for $Z' \to e^+ e^-$ is given by
\begin{align}
\Gamma(Z' \to e^+ e^-) &=\frac{Q_f^2 \epsilon^2 e^2 c_W^2}{12 \pi} m_{Z'} \left(1+ {2 m_e^2 \over m_{Z'}^2}\right) \left(1- {4 m_e^2 \over m_{Z'}^2}\right)^{1/2} \nonumber\\
& \approx 1.87 \times 10^{-11} \, {\rm GeV} \left(\epsilon \over 10^{-4} \right)^2 \left(m_{Z'} \over 1 \, {\rm GeV} \right)^2.
\end{align}
Its lifetime is much shorter than 1 sec in the parameter space of our interest.

The light $\phi$ and/or $Z'$ may contribute to the effective neutrino number $N_{\rm eff}$, which is another possible constraint in the model.
Since the mediators $Z'$ and $\phi$ decay before 1 sec, there is no relativistic extra degree freedom which mimics the neutrino at 
the recombination era ($T_{\rm CMB} \approx$ 4 eV).
Another potential source for $\Delta N_{\rm eff}$ is light mediator with mass below 1 MeV.
It mainly decays into $e^\pm$ or $\gamma$ not into $\nu$,  making the difference between the temperatures $T_\gamma$ and $T_\nu$ larger 
than the one given by the standard cosmology by imparting its entropy only to 
$\gamma$~\cite{Matsumoto:2018acr}. 
This also causes $\Delta N_{\rm eff} \not= 0$.
We evade this problem by taking their masses greater than 1 MeV.

\begin{figure}
\includegraphics[width=0.40\linewidth]{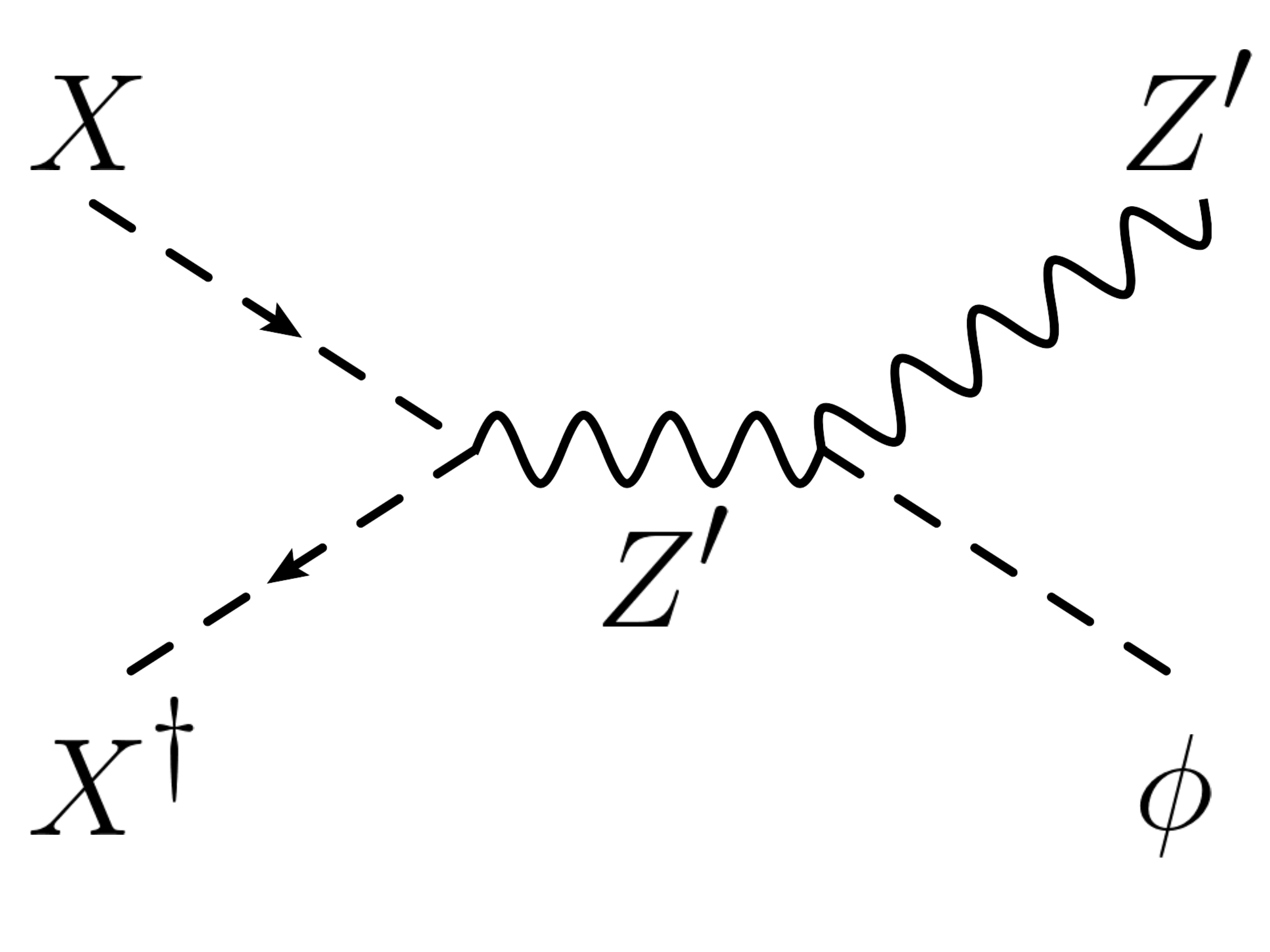}
\includegraphics[width=0.50\linewidth]{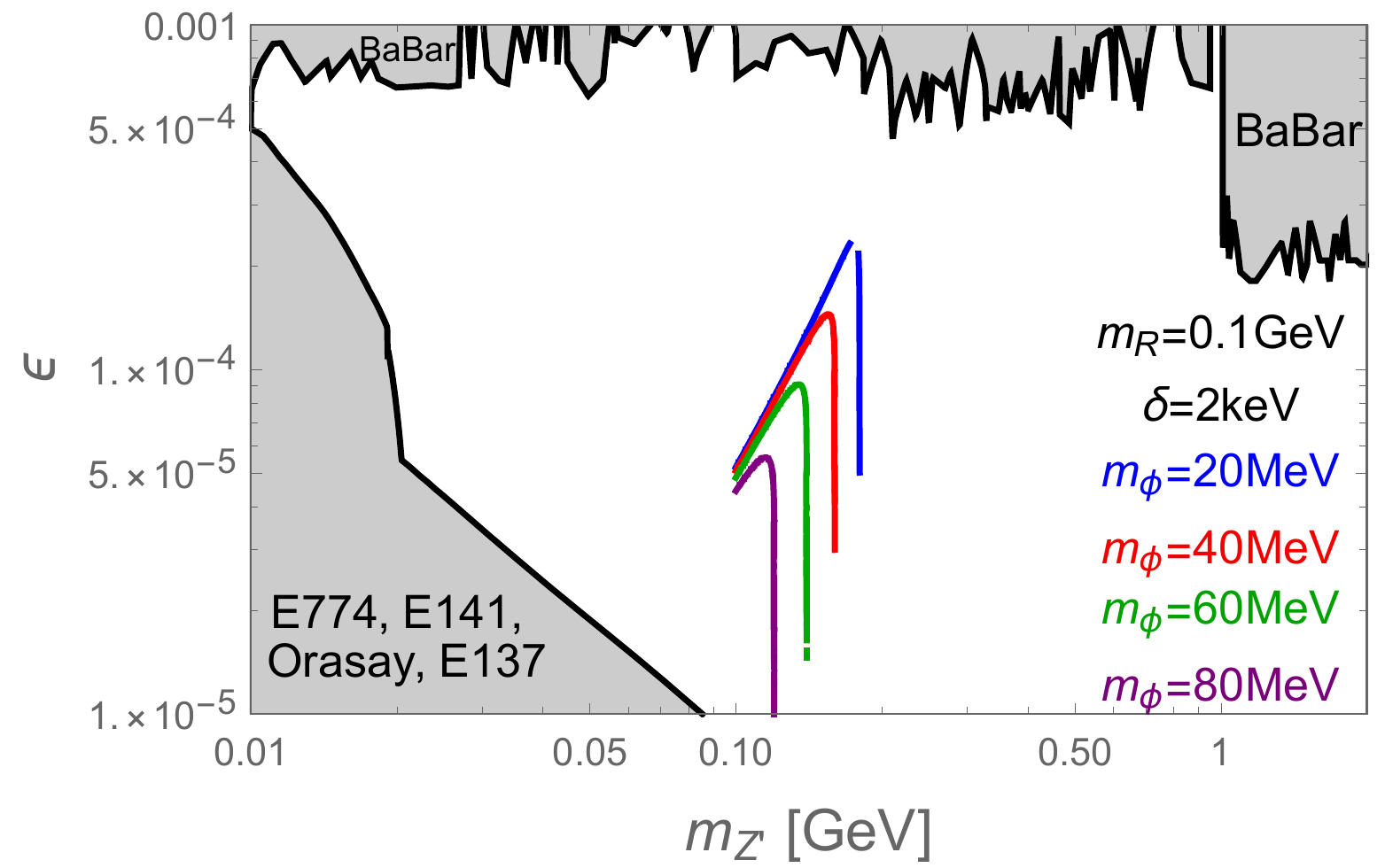}
\caption{ \label{SDM} 
({\it left}) Feynman diagrams relevant for thermal relic density of DM: 
 $X X^\dagger \rightarrow Z' \phi$ and ({\it right}) the region in the $( m_{Z'}, \epsilon)$ plane 
that is allowed for the XENON1T electron recoil excess and the correct thermal relic density 
for scalar DM case for $\delta = 2$ keV : (a) $m_{\rm DM} = 0.1$ GeV. Different colors represents 
$m_\phi = 20,40,60,80$ MeV.    The gray areas are excluded by various experiments, from BaBar \cite{Lees:2014xha}, E774 \cite{Bross:1989mp}, E141 \cite{Riordan:1987aw}, Orasay \cite{Davier:1989wz}, and E137 \cite{Batell:2014mga}, assuming $Z' \rightarrow X_R X_I$ is kinematically forbidden.
}
\end{figure}

In this restricted region of parameter space the main channel for the relic density is 
$X X^\dagger \to Z^{\prime *}\to Z' \phi$ (see the left panel in Fig. \ref{SDM} ) 
\footnote{From now on we call $H_2$ as $\phi$.}. 
The leading contribution to cross section is $p$-wave with the cross section
\begin{align}
\sigma v & \simeq \frac{g_X^4 v^2}{384  \pi \,m_X^4 (4 m_X^2-m_{Z'}^2 )^2}\left(16 m_X^4 + m_{Z'}^4 + m_\phi^4+40 m_X^2 m_{Z'}^2 -8 m_X^2 m_\phi^2 -2 m_{Z'}^2 m_\phi^2\right) \nonumber\\
&\times \Big[\left\{4 m_X^2-(m_{Z'}+m_\phi)^2\right\}\left\{4 m_X^2-(m_{Z'}-m_\phi)^2\right\}\Big]^{1/2}+{\cal O}(v^4),
\end{align}
where $m_\phi  \equiv \sqrt{2 \lambda_\phi} v_\phi \simeq m_{H_2}$.
The resulting dark matter density is obtained by~\cite{Griest:1990kh}
\begin{align}
\Omega_X h^2 = \frac{2 \times 8.77 \times 10^{-11} \, {\rm GeV}^{-2} \,x_f}{g_*^{1/2} (a+3 b/x_f)},
\label{eq:Omega}
\end{align}
where $a$ and $b$ are defined by $\sigma v = a + b v^2$, and the additional factor 2 
comes from the fact that $X$ is a complex scalar instead of real scalar.
For example, with $m_X =1$ GeV, $m_{Z'} =1.2$ GeV, $m_\phi=0.2$ GeV, $\lambda_\phi = 4.5 \times 10^{-5}$, 
 $g_* \approx 10$, and $x_f \approx 10$, we can explain the current DM relic abundance: $\Omega_X h^2 \approx 0.12$.
Other $p$-wave contributions include $X X^\dagger \to Z^{\prime *} \to f \bar{f}$ where $f$ is an SM fermion.
But these contributions are suppressed by $\epsilon^2$ compared to the above annihilation, and we neglect them.
The SM $Z$ boson contributions are further suppressed by both small mixing angle $\epsilon^2$ and small mass ratio $m_{Z'}^4/m_{Z}^4$.
 
To calculate the inelastic down-scattering cross section for the XENON1T anomaly, instead of $X$ and $X^\dagger$ we now consider
two real scalars $X_R, X_I$ with mass difference $\delta$. With the kinetic mixing term given in (\ref{eq:model}) we get the dark-gauge interactions 
with the DM and the electron~\cite{Babu:1997st}
\begin{align} 
{\cal L} \supset g_X Z^{\prime\mu} (X_R \partial_\mu X_I - X_I \partial_\mu X_R) - \epsilon\, e c_W Z^\prime_\mu \overline{e} \gamma^\mu e,
\end{align} 
where $c_W$ is the cosine of the Weinberg angle,  $Z$ and $Z'$ are mass eigenstates, and we assumed that $\epsilon(\sim 10^{-4})$ is small.
The cross section for the inelastic scattering $X_R e \to X_I e$ for $m_X \gg m_e$ and small momentum transfer is given by
\begin{align}
\sigma_e = \frac{16 \pi \epsilon^2 \alpha_{\rm em} \alpha_X c_W^2  m_e^2}{m_{Z'}^4},
\label{eq:sigma_e}
\end{align}
where $\alpha_{\rm em}\simeq 1/137$ is the fine structure constant and $\alpha_X \equiv g_X^2/4 \pi$.
This can be used to predict the differential cross section of the dark matter scattering off the xenon atom for the DM velocity $v$, which reads
\begin{align}
\frac{d \sigma v}{d E_R}=\frac{\sigma_e}{2 m_e v} \int_{q_-}^{q_+} a_0^2 q dq K(E_R,q),
\end{align}
where $E_R$ is the recoil energy, $q$ is the momentum transfer, $K(E_R,q)$ is the atomic excitation factor.
From energy conservation we obtain the relation~\cite{Harigaya:2020ckz},
\begin{align}
E_R = \delta + v q \cos\theta -\frac{q^2}{2 m_R},
\end{align}
where $\theta$ is the angle between the incoming $X_R$ and the momentum transfer $\boldsymbol{q}=\boldsymbol{p'}_e -\boldsymbol{p}_e$.
The integration limits are~\cite{Harigaya:2020ckz},
\begin{align}
q_\pm &\simeq m_R v \pm \sqrt{m_R^2 v^2 - 2 m_R (E_R -\delta)}, \quad \text{for} \; E_R \ge \delta, \nonumber\\
q_\pm &\simeq \pm m_R v +\sqrt{m_R^2 v^2 - 2 m_R (E_R -\delta)}, \quad \text{for} \; E_R \le \delta.
\end{align}
Then we can obtain the differential event rate for the inelastic scattering of DM with electrons in the xenon atoms given by
\begin{align}
\frac{d R}{d E_R} = n_T {n_R} \frac{d \sigma v}{d E_R},
\end{align}
where $n_T \approx 4 \times 10^{27}/$ton is the number density of xenon atoms and $n_R \approx 0.15 \, {\rm GeV}/m_R/{\rm cm}^3$ is
the number density of the heavier DM component $X_R$, assuming $n_R =n_I$. Integrating over $E_R$, we get the event rate
\begin{align}
R \approx 3.69 \times 10^9 \, \epsilon^2 \, g_X^2 \left( 1 {\rm GeV} \over m_R \right) \left( 1 {\rm GeV} \over m_{Z'} \right)^4 /\text{ton}/\text{year}.
\end{align}

Since $X_R$ is a dark matter component in our model with the same abundance with $X_I$, its lifetime should be much longer than the age of the universe.
It can decay via $X_R \to X_I \gamma \gamma \gamma$ as shown in~\cite{Harigaya:2020ckz}. 
Its decay into three-body final state, $X_R \to X_I \nu \overline{\nu}$, is also possible in our model. The relevant interactions are
\begin{align}
{\cal L} \supset \epsilon g_X s_W Z^\mu (X_R \partial_\mu X_I -X_I \partial_\mu X_R) -{g_Z \over 2} Z_\mu \overline{\nu}_L \gamma^\mu \nu_L.
\end{align}
The decay width is given by
\begin{align}
\Gamma \simeq \frac{\epsilon^2 \alpha_X s_W^2}{5\sqrt{2} \pi^2} \; \frac{G_F \delta^5 }{m_Z^2} \simeq 1.9 \times 10^{-49} \,\text{GeV}
\left(\epsilon \over 10^{-4}\right)^2 \left(\alpha_X \over 0.078\right)\left(\delta \over 2\,\text{keV}\right)^5.
\label{eq:GammaX}
\end{align}
Although this channel is much more effective than $X_R \to X_I \gamma\gamma\gamma$ considered in~\cite{Harigaya:2020ckz},
the lifetime of $X_R$ is still much longer than the age of the universe.

In the right panel of Fig. \ref{SDM} , we show the allowed region in the $(m_{Z'}, \epsilon)$ plane 
where we can explain the XENON1T excess with correct thermal relic density of DM within the 
standard freeze-out scenario.  For illustration, we chose the DM mass to be $m_R = 0.1$ GeV, 
and varied the dark Higgs mass $m_\phi = 20, 40, 60, 80$ MeV denoted with different colors. 
The sharp drops on the right allowed region is from the kinematic boundary, 
$m_{Z'} + m_\phi < 2 m_R$. It is nontrivial that we could explain the XENON1T excess with 
inelastic DM models with spontaneously broken  $U(1)_X \rightarrow Z_2$ gauge symmetry. 
In particular it is important to include light dark Higgs for this explanation. 
It would be straightfoward to scan over all the parameters to get the whole allowed region.

\subsection{Fermion DM model }

We start from a dark $U(1)$ model, with a Dirac fermion dark matter (DM) $\chi$ appointed with a nonzero dark $U(1)$ charge $Q_\chi$ and dark photon.   We also introduce a complex dark Higgs field $\phi$, which takes a nonzero vacuum expectation value, generating nonzero mass for the dark photon. We shall consider a special case where $\phi$ breaks the dark $U(1)$ symmetry into a dark $Z_2$ symmetry with a judicious choice of its dark charge $Q_\phi$. 

Then the gauge invariant and renormalizable Lagrangian for this system is given by 
\begin{eqnarray}
\mathcal{L} & = & -\frac{1}{4} \hat{X}^{\mu \nu} \hat{X}_{\mu \nu} 
- \frac{1}{2} \sin \epsilon \hat{X}_{\mu\nu}B^{\mu\nu} 
+ \overline{\chi} \left( i\slashed{D} - m_{\chi} \right) \chi + D_{\mu} \phi^\dagger D^{\mu} \phi 
\label{Lag_fermion}
\\
& - & \mu^2 \phi^\dagger  \phi - \lambda_{\phi} |\phi|^4 
-  \frac{1}{\sqrt{2}} \left( y \phi^\dagger \overline{\chi^C} \chi + \text{h.c.} \right) 
- \lambda_{\phi H} \phi^\dagger \phi H^\dagger H
\nonumber 
\end{eqnarray}
where $\hat{X}_{\mu \nu} = \partial_{\mu} \hat{X}_{\nu} - \partial_{\nu} \hat{X}_{\mu}$. 
$D_{\mu} = \partial_{\mu} + i g_X Q_X \hat{X}_{\mu}$ is the covariant derivative, 
where $g_X$ is the dark coupling constant, and $Q_X$ denotes the dark charge of $\phi$ and $\chi$: 
$Q_{\phi}  = 2,  Q_{\chi} = 1$, respectively.   
Then $U(1)_X$ dark gauge symmetry is spontaneously broken into its $Z_2$ subgroup, and 
the Dirac DM $\chi$ is split into two Majorana DM $\chi_R$ and $\chi_I$ defined as 
\begin{eqnarray}
\chi &  = & \frac{1}{\sqrt{2}} ( \chi_R + i \chi_I ) ,
\\
\chi^c &  = & \frac{1}{\sqrt{2}} ( \chi_R - i \chi_I ) ,
\\
\chi_R^c & = & \chi_R , \ \ \  \chi_I^c = \chi_I , 
\end{eqnarray}
with 
\begin{equation}
m_{R,I} = m_\chi \pm  y v_\phi = m_\chi \pm \frac{1}{2} \delta .
\end{equation} 

We assume  $y>0$ so that $\delta \equiv m_R - m_I = 2 y v_\phi > 0$.    
Then the above Lagrangian is written as 
\begin{eqnarray}
{\cal L} & = & \frac{1}{2} \sum_{i=R,I} \overline{\chi_i} \left( i \slashed{\partial} - m_i \right) \chi_i 
- i \frac{g_X}{2}  ( Z'_\mu+\epsilon s_W Z_\mu) \left( \overline{\chi_R} \gamma^\mu \chi_I 
- \overline{\chi_I} \gamma^\mu \chi_R \right)  
\\
& - & {1 \over 2} y h_\phi \left( \overline{\chi_R} \chi_R  - \overline{\chi_I} \chi_I \right),
\end{eqnarray}
where $h_\phi$ is neutral CP-even component of $\phi$ as defined in (\ref{eq:ScalarDecomp}).

\begin{figure}
\includegraphics[width=0.40\linewidth]{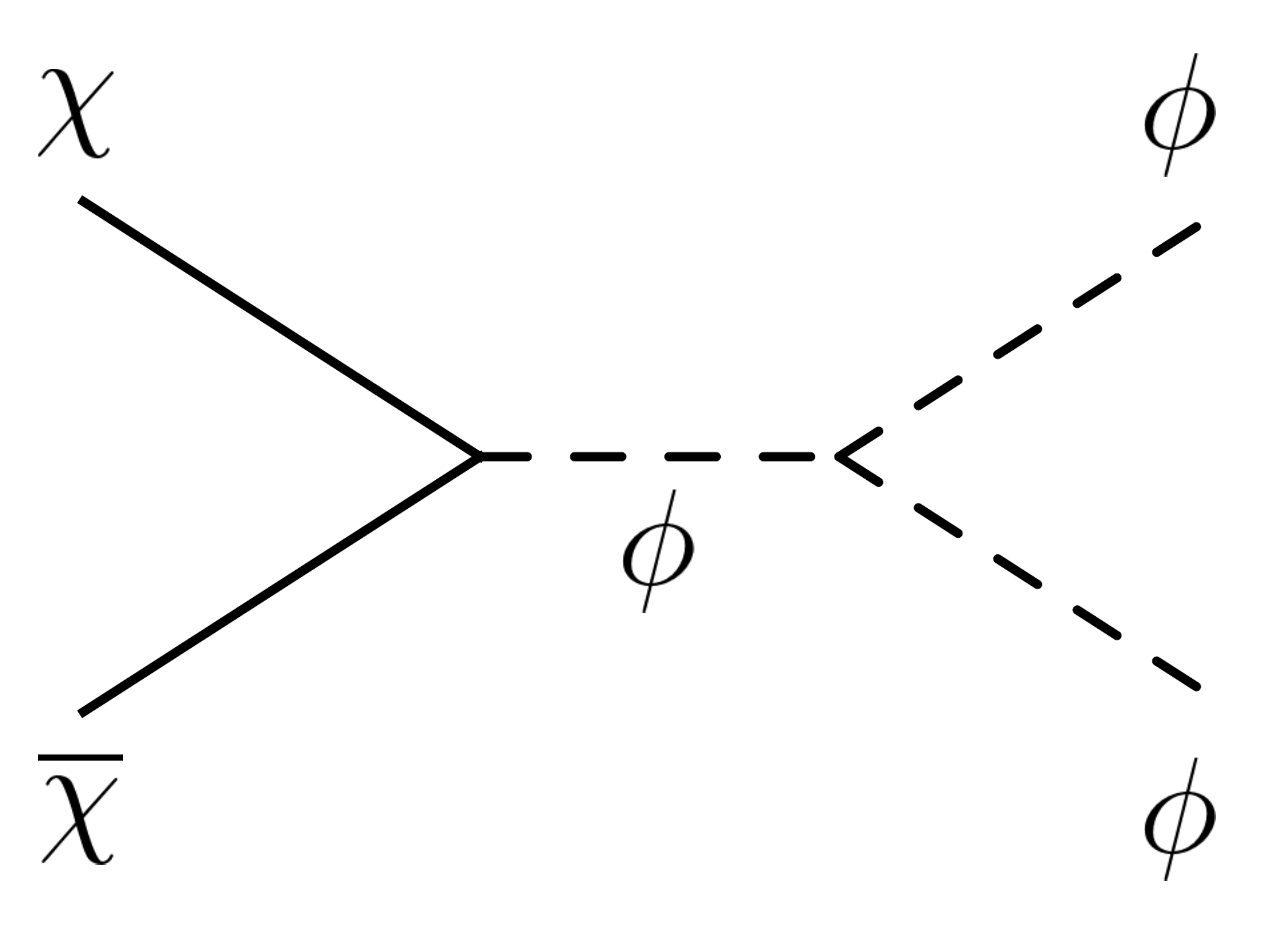}
\hspace{0.5cm}
\includegraphics[width=0.40\linewidth]{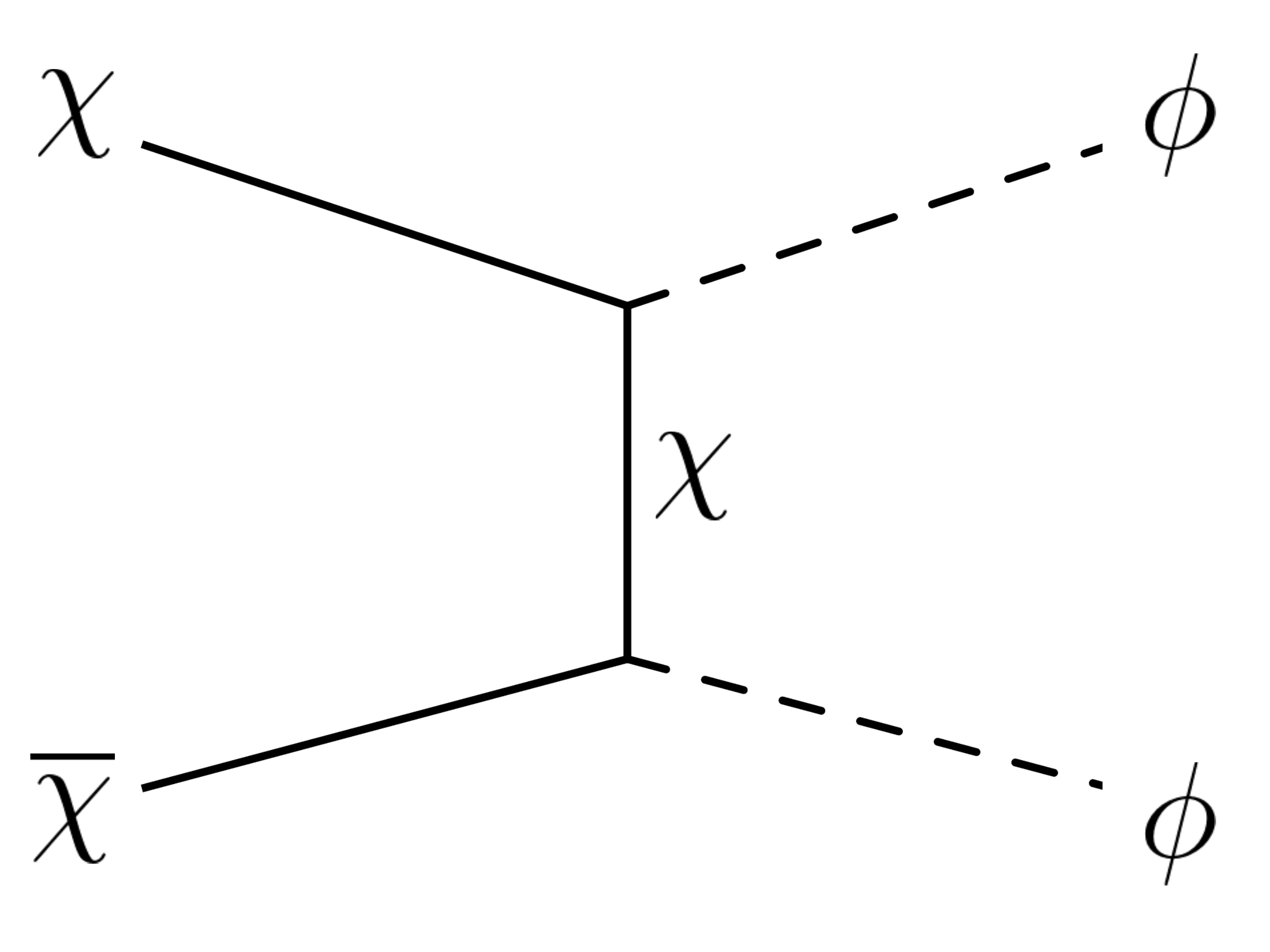}
\\
\vspace{0.5cm}
\includegraphics[width=0.50\linewidth]{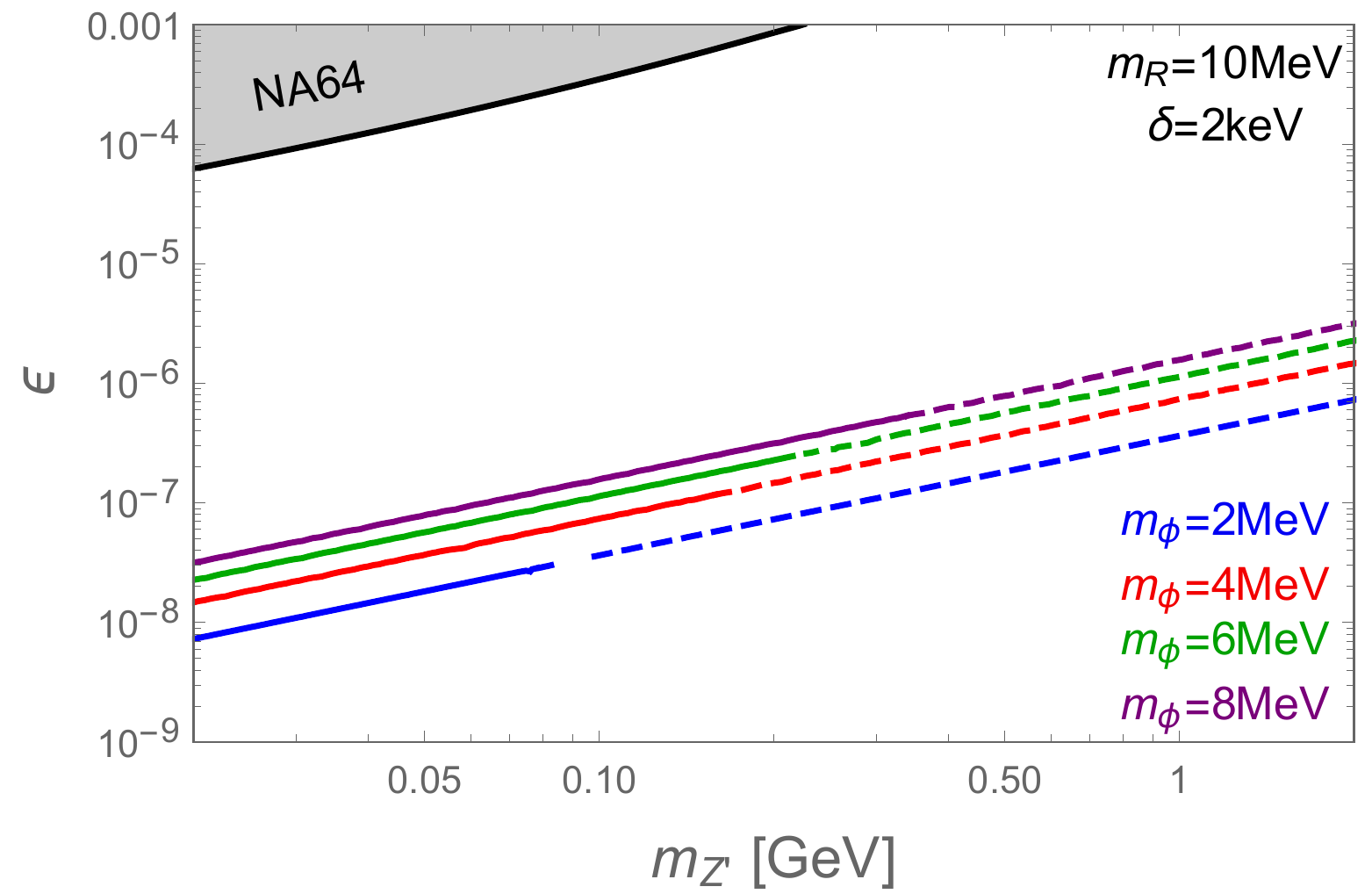}
\caption{ \label{FDM} 
({\it top}) Feyman diagrams for $\chi \bar{\chi} \rightarrow \phi\phi$. 
({\it bottom}) the region in the $( m_{Z'}, \epsilon)$ plane 
that is allowed for the XENON1T electron recoil excess and the correct thermal relic density 
for fermion DM case for $\delta = 2$ keV and the fermion DM mass to be $m_R = 10$ MeV. 
Different colors represents $m_\phi = 2, 4, 6, 8$ MeV.    The gray areas are excluded by various 
experiments, assuming $Z' \rightarrow \chi_R \chi_I$ is kinematically allowed, and the experimental 
constraint is weaker in the $\epsilon$ we are interested in, compared with the scalar DM case 
in Fig. 1 (right). We also show the current experimental bounds by NA64 \cite{NA64:2019imj}. 
}
\end{figure}

When we calculate the DM relic density, we can assume the mass difference is small compared 
to the DM mass as in the case of the scalar DM, {\it i.e.} $m_\chi \simeq m_R \simeq m_I$.
For the fermionic DM the annihilation processes into the scalar pair, $\stackrel{(-)}{\chi} \stackrel{(-)}\chi \to \phi \phi$,  are $p$-wave.
To evade the CMB constraint we suppress the $s$-wave annihilation by assuming $2 m_\chi<2m_{Z'}, m_{Z'}+m_\phi$.
The calculation of the annihilation process (the top panel of Fig. \ref{FDM}) yields
\begin{align}
\sigma v = \frac{y^2 v^2 \, \sqrt{m_\chi^2-m_\phi^2}}{96\pi m_\chi} 
\Bigg[\frac{27 \lambda_\phi^2 v_\phi^2}{(4 m_\chi^2 -m_\phi^2)^2}
+\frac{4 y^2 m_\chi^2 (9 m_\chi^4 -8 m_\chi^2 m_\phi^2 +2 m_\phi^4)}{(2 m_\chi^2 -m_\phi^2)^4}
 \Bigg] +{\cal O}(v^4),
\end{align}
where $m_\phi  \equiv \sqrt{2 \lambda_\phi} v_\phi \simeq m_{H_2}$.
The current DM relic abundance is obtained by (\ref{eq:Omega}).  Since the annihilation cross 
section for the fermion DM case if proportional to $y^2 \propto (\delta/v_\phi )^2$ and 
$\delta \sim 2$ keV, the $v_\phi$  should be not too large. If we ignore $m_\phi$ in the above 
equation, we find that $m_\chi \sim O(1-10)$ MeV will be required to get the correct thermal relic 
density, and dark Higgs $\phi$ should be even lighter. Therefore dark sector particles in this
case should be lighter than the scalar DM case. 
Later for illustration, we will consider $m_\chi \sim O(10)$ MeV. and $m_\phi \sim O(1)$ MeV 
to get the correct DM relic density and explain the XENON1T excess.

Now let's consider the inelastic scattering of DM  with the electron in the xenon atom to explain the XENON1T anomaly.
The scattering occurs through the interactions
\begin{align}
{\cal L} \supset 
- i \frac{g_X}{2}   Z'_\mu \left( \overline{\chi_R} \gamma^\mu \chi_I 
- \overline{\chi_I} \gamma^\mu \chi_R \right)  
- \epsilon\, e c_W Z^\prime_\mu \overline{e} \gamma^\mu e,
\end{align}
It turns out that in the limit $m_\chi \gg m_e$, $\sigma_e$ has exactly the same form with (\ref{eq:sigma_e}) of the scalar DM case.

We require the $\chi_R$ to be long-lived so that it is also a main component of the dark matter.
It decays mainly via the SM $Z$-mediating $\chi_R \to \chi_I \nu \overline{\nu}$, using the interactions
\begin{align}
{\cal L} \supset -{i \over 2} \epsilon s_W g_X Z_\mu (\overline{\chi}_R \gamma^\mu \chi_I -\overline{\chi}_I \gamma^\mu \chi_R)
-{1 \over 2} g_Z Z_\mu \overline{\nu}_L \gamma^\mu \nu_L.
\end{align}
The expression for the decay with, $\Gamma(\chi_R \to \chi_I \nu \overline{\nu})$, also agrees exactly with (\ref{eq:GammaX}).
As shown in (\ref{eq:GammaX}), the lifetime of $\chi_R$ is much longer than the age of the universe, which guarantees the $\chi_R$ is as good a dark matter as $\chi_I$.

In the bottom panel of Fig. \ref{FDM}, we show the allowed region in the $(m_{Z'}, \epsilon)$ plane 
where we can explain the XENON1T excess with correct DM thermal relic density within the 
standard freeze-out scenario.  For illustration, we choose the fermion DM mass to be 
$m_\chi = m_R = 10$ MeV,   and varied the dark Higgs mass $m_\phi = 2, 4, 6, 8$ MeV denoted 
with different colors.    Note that the kinetic mixing $\epsilon \sim 10^{-7 \pm 1}$, which is much 
smaller than the scalar DM case. 
We have checked if the gauge coupling $g_X$ and the quartic coupling of dark Higgs 
($\lambda_\phi$) remain in the perturbative regime. The solid (dashed) lines denote the region where 
$g_X$ satisfy (violate) perturbativity condition, depending $\alpha_X < 1$ or not.    
Within this allowed region, $\lambda_\phi$ remain perturbative. 
Again it is nontrivial that we could explain the XENON1T excess with inelastic fermion DM 
models with spontaneously broken  $U(1)_X \rightarrow Z_2$ gauge symmetry. 
In particular it is important to include light dark Higgs for this explanation as in the scalar DM case.

\section{Conclusion}

In this paper, we showed that the electron recoil excess reported by XENON1T Collaboration could
be accounted for by exothermic DM scattering on atomic electron in Xe, with sub-GeV light 
DM: $m_X \sim O(100)$ MeV for the scalar and $m_\chi \sim O(10)$ MeV for the fermion DM, 
and dark Higgs $\phi$ neing even lighter that DM particle for both cases. 
Dark photon should be heavier than DM in order that we can forbid the DM pair  annihilation 
into the $Z' Z'$ channels. 
This scenario could be described by DM models with dark $U(1)$ gauge symmetry broken 
into its $Z_2$ subgroup by Krauss-Wilczek mechanism. And dark photon $Z'$ and dark Higgs $\phi$ 
in such dark gauge models play important roles in DM phenomenology.   
In particular in the calculation of thermal relic density,  new channels involving a dark Higgs can 
open $X X^\dagger \rightarrow \phi Z'$ and  $\chi \bar{\chi} \rightarrow \phi \phi$, 
which are $p$-wave annihilations. Then one could evade  the stringent constraints from CMB 
for such light DM.   Other dangerous $s$-channel annihilations can be kinematically forbidden 
by suitable choice of parameters. 
Thus the exothermic scattering in inelastic $Z_2$ DM models within standard freeze-out scenario 
can explain the XENON1T excess without modifying early universe cosmology. 
We emphasize again that the existence of dark Higgs $\phi$ is crucial for us to get the desired DM 
phenomenology to explain the XENON1T excess with the correct thermal relic density in case of 
both  scalar and fermion DM models within the standard freeze-out scenario.

\section*{Note Added}
While we were preparing this manuscript, there appeared a few papers which explain
the XENON1T excess in terms of scalar or fermion exothermic DM \cite{Harigaya:2020ckz,Su:2020zny,Lee:2020wmh,Bramante:2020zos,Baryakhtar:2020rwy,An:2020tcg}. 
Our paper is different from these previous works in that we consider dark $U(1)$ gauge symmetry 
broken to its $Z_2$ subgroup by dark Higgs mechanism, and include the light dark Higgs in the 
calculations of thermal relic density for the two component DM in the standard freeze-out scenario:
$X X^\dagger \rightarrow Z' \phi$ for scalar DM and $\chi \overline{\chi} \rightarrow \phi\phi$.
Other $s$-wave channels are kinematically forbidden by suitable choice of mass parameters. 
This possibility of dark Higgs in the final state in the (co)annihilation channels are not included 
in other works.  By including these new channels, we could achieve the correct thermal relic
density and the desired heavier DM fluxes on the Xe targets simultaneously, without conflict with
strong constraints on light DM annihilations from CMB.
And the models considered in this paper is renormalizable and DM stability is guaranteed by 
underlying $U(1)$ dark gauge symmetry and its unbroken $Z_2$  subgroup.

\begin{acknowledgments}
The work is supported in part by KIAS Individual Grants, Grant No. PG021403 (PK) and 
Grant No. PG074201 (JK) at Korea Institute for Advanced Study, and 
by National Research Foundation of Korea (NRF) Grant No. NRF-2018R1A2A3075605 (SB) 
and No. NRF-2019R1A2C3005009 (PK), funded by the Korea government (MSIT). 
\end{acknowledgments}

\bibliographystyle{utphys}

\bibliography{literature}

\end{document}